\def\year{2022}\relax
\documentclass[letterpaper]{article} 
\usepackage{aaai22}  
\usepackage{times}  
\usepackage{helvet}  
\usepackage{courier}  
\usepackage[hyphens]{url}  
\usepackage{graphicx} 
\usepackage{natbib}  
\usepackage{caption} 
\DeclareCaptionStyle{ruled}{labelfont=normalfont,labelsep=colon,strut=off} 
\usepackage{algorithm}
\usepackage{algorithmic}

\usepackage{newfloat}
\usepackage{listings}
\floatstyle{ruled}
\newfloat{listing}{tb}{lst}{}
\floatname{listing}{Listing}
\title{mask-Net: Learning Context Aware Invariant Features using Adversarial Forgetting (Student Abstract)}
\author{
    Hemant Yadav\textsuperscript{\rm 1}, Atul Anshuman Singh\textsuperscript{\rm 1}\equalcontrib, Rachit Mittal\textsuperscript{\rm 1}\equalcontrib, \\
    Sunayana Sitaram\textsuperscript{\rm 2}, Yi Yu\textsuperscript{\rm 3}, Rajiv Ratn Shah\textsuperscript{\rm 1}
}
\affiliations{
    \textsuperscript{\rm 1}MIDAS, IIIT Delhi, India, 
    \textsuperscript{\rm 2}Microsoft Research India 
    \textsuperscript{\rm 3}National Institute of Informatics, Tokyo \\
    \{hemantya, rachit18302, rajivratn\}@iiitd.ac.in, atula.ec@nsit.net.in, sunayana.sitaram@microsoft.com, yiyu@nii.ac.jp,
}

\usepackage{bibentry}

\begin{document}

\maketitle

\begin{abstract}
Training a robust system, \emph{e.g.,} Speech to Text (STT), requires large datasets. Variability present in the dataset such as unwanted nuisances and biases are the reason for the need of large datasets to learn general representations. In this work, we propose a novel approach to induce invariance using adversarial forgetting (AF). Our initial experiments on learning invariant features such as accent on the STT task achieve better generalizations in terms of word error rate (WER) compared to the traditional models. We observe an absolute improvement of 2.2\% and 1.3\%  on out-of-distribution and in-distribution test sets, respectively.

\end{abstract}

\begin{figure*}
 \centering
 \scalebox{0.3}[0.25]{
    \centering
     \includegraphics{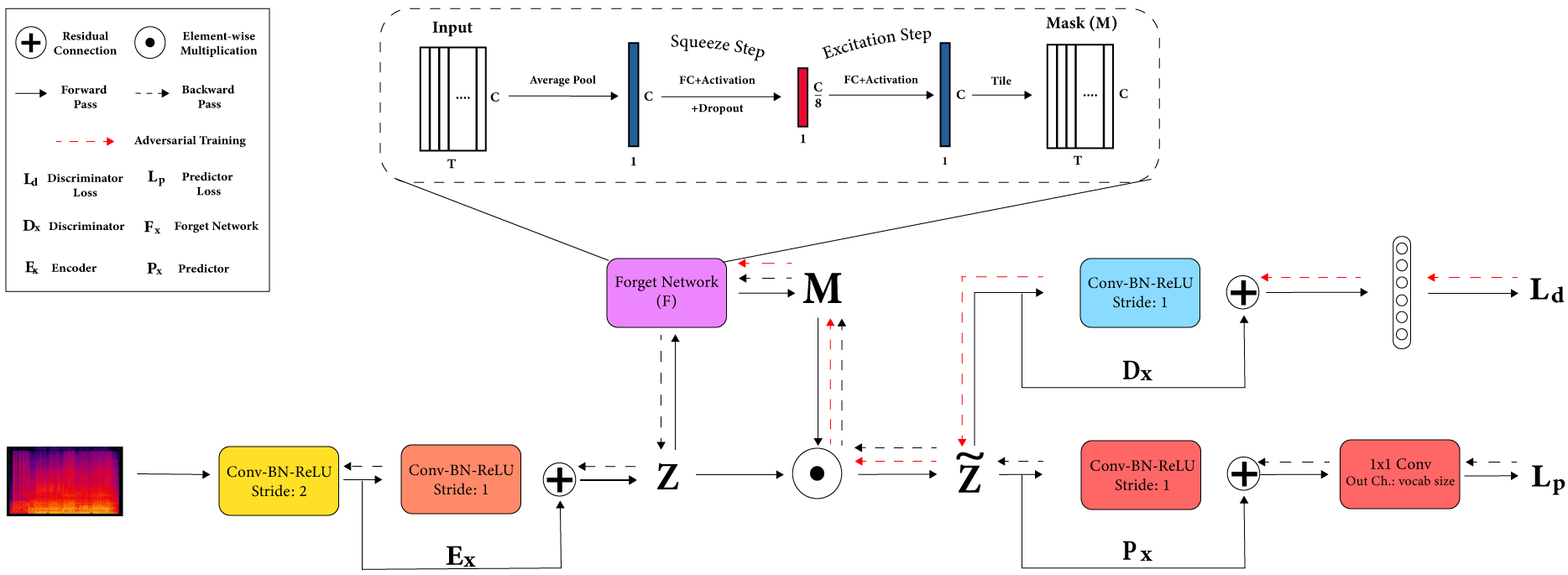}}
     \caption{mask-Net architecture for learning invariant features using adversarial forgetting. The rectangular blocks represent 1D CNN layers with Batch-Norm (BN) and ReLU and $x$ denotes the number of repeated blocks as proposed by the authors of Jasper paper \cite{li2019jasper}. Each block can have $n$ layers internally. Residual connections are only active if the number of layer within the block are greater than 1. Dotted arrows show the flow of gradients corresponding to each loss function.}
     \label{fig:AF}
\end{figure*}

\section{Introduction}\label{introduction}

Given a task with input $X$ to predict a set of labels $Y$, deep learning (DL) systems often learn an undesired dependence on unwanted factors present in the input given the task \cite{domingos2012few} \emph{e.g.,} STT. Therefore the need arises to learn invariant features to avoid the undesired dependencies. Previous work in this direction includes enforcing the invariance throughout all the layers of the encoder network using gradient reversal learning (GRL) technique \cite{ganin2015unsupervised}. \cite{jaiswal2020invariant} propose to model the problem of learning invariant features by learning a mask $M \in (0,1)$, given $X$, which on element wise multiplication with the encoder output $Z$ produces an invariant feature $\tilde Z$. Our proposed method builds upon the two major shortcomings of this work.

Firstly, in their model setup, the authors learn low level features twice, once in the encoder and then again in the forget net module. It results in learning redundant features and increased training time since both the modules have an identical architecture. Secondly, if the main task is based on time series data, \emph{e.g.,} STT or machine translation, their is no notion of global context when learning the mask \emph{M}. This hampers learning what to retain and what to forget \emph{i.e.,} the $M$ values. Furthermore, each value of mask $M$ is learnt independently such that there is no correlation between the mask values over different time steps. This results in different mask values for different time steps, which is an undesirable behaviour.

In this work, we propose a novel approach to address these issues. Our method is build upon the paper \cite{jaiswal2020invariant} with details provided in the next section.

\section{Method}\label{method}

We propose a novel approach to induce invariance to unwanted factors, \emph{e.g.,} accent present in the audio for the STT task. Let $X$ be the input, $Y$ be the output and, $A$ be the unwanted bias. We propose an architecture with one encoder module \emph{E} with output $Z$, one forget net module \emph{F} with output $M \in (0,1)$, one discriminator module \emph{D} with output $A$ and, one predictor module \emph{P} with output $Y$. The mask $M$ is then multiplied element wise with $Z$ to get $\tilde Z$ as shown in Figure \ref{fig:AF}.

As shown in Figure \ref{fig:AF} the input to forget net is $Z$ instead of the raw input $X$. Therefore we bypass the redundant step to learning the low level features through the forget net module leading to reduction in training time \footnote{For details refer to the supplementary material.}. Secondly, to introduce the notion of global context we propose to create the forget net module similar to the squeeze and excitation module presented in ContextNet \cite{han2020contextnet}. As shown in Figure \ref{fig:AF} we first average pool over all the time steps to introduce the global context and then apply a linear layer such that the output dimension is reduced by eight (squeeze step). In the next step we again apply a linear layer such that the output dimension is restored to the previous step (excitation step). In the last step we tile the output to match the dimension of $Z$. The last step makes sure that all the time steps have the same mask value and thus the embeddings at each time step to map to a similar feature. We propose the above two architectural changes compared to the \cite{jaiswal2020invariant} and call the resulting network mask-Net (M-Net) such that it learns context aware invariant features using adversarial forgetting as shown in Figure \ref{fig:AF}.

The mask $M$ is learned in an adversarial manner similar to the GRL technique with a distinction that the gradients from the discriminator only flows up to the forget net as shown by the dotted red lines in Figure \ref{fig:AF}. Therefore the encoder learning is purely driven by the main task. This is very useful in case we want to use the $Z$ for some other downstream task which requires the accent information. A practical application is to learn a common strong encoder using self supervised learning (SSL) and then use our method to learn $n$ masks given $n$ biases. This is not possible in the standard GRL technique. Lastly we expect our method to achieve results comparable to the standard GRL training technique if not better.

\section{Preliminary Evaluation and Results}

Similar to the \cite{jain2018improved}, we evaluate our approach on a subset of Common Voice V1\footnote{https://commonvoice.mozilla.org/en/datasets} dataset, 34 hrs for training. Where accent classification is the discriminator task and STT is the predictor/main task. No data augmentation has been applied on the input spectrogram for training and testing purposes. Predictor and discriminator are trained with connectionist temporal classification \cite{graves2006connectionist} and cross entropy loss respectively. We report WER with a beam width of 512.  All models are trained using mixed precision with Pytorch\footnote{Pre-trained models, code and training configurations will be made available on GitHub upon acceptance.}.

Surprisingly our Japser \cite{li2019jasper} based baseline achieves far better WERs when compared to one of the recent work by \cite{jain2018improved}, in which the authors use utterance-level accent embeddings. This goes on to show that the baseline we provide is very strong. Therefore we only make comparison to our baseline and try to improve on that. We observe an overall improvement across all the test cases when compared with baseline as shown in Table \ref{table:paper comparision}.

Our proposed method achieves comparable results to standard GRL technique but achieves lower classification accuracy on discriminator, which is desirable for achieving true invariant features \cite{jaiswal2020invariant}. We attribute this chaotic behaviour of the metrics to the small training dataset size, 34 hrs. Therefore further experiments are needed to be done on large datasets with varying biases to report a general trend.

\begin{table}[h]
\centering
    \scalebox{0.75}{
    \begin{tabular}{|c|c|c|c|c|}
    \hline
      \textbf{\%} & \textbf{Test} & \textbf{TestIN} & \textbf{TestNZ}  & \textbf{Accuracy} \\
      \hline
     \cite{jain2018improved} & 19.74 & 51.2 & 22.7 & -- \\
      \hline
      Baseline \cite{li2019jasper} & 7.7      & 33.6     & 6.4 & --\\
     \hline
     Multi-task    & 6.8    & 32.9     & 5.4 & --\\
     \hline
     Gradient-reversal    & 6.2    & 30.3     & 4.9 & 46.7\\
     \hline
     \textbf{Ours}  & \textbf{6.4}  & \textbf{30.5}   & \textbf{5.1} & \textbf{41}\\
     \hline
        \end{tabular}}
\caption{WER is reported on in-distribution test set \emph{i.e.,} Test and out-of-distribution test sets \emph{i.e.,} TestIN (India) and TestNZ (New Zealand). Similarly to \cite{jaiswal2020invariant}, lower accuracy is desirable.}
\label{table:paper comparision}
\end{table}

\section{Conclusion and Future Work}

In this work we proposed a novel approach to learn invariant representations with the added benefits of global context aware mask, faster training time and, uniform mask values over all the time steps. Future work includes running experiments on bigger dataset with different biases. Furthermore to achieve true global context, we would try attention based forget net module to learn the mask $M$.


\bibliography{aaai22}

\begin{thebibliography}{7}
\providecommand{\natexlab}[1]{#1}

\bibitem[{Domingos(2012)}]{domingos2012few}
Domingos, P. 2012.
\newblock A few useful things to know about machine learning.
\newblock \emph{Communications of the ACM}, 55(10): 78--87.

\bibitem[{Ganin and Lempitsky(2015)}]{ganin2015unsupervised}
Ganin, Y.; and Lempitsky, V. 2015.
\newblock Unsupervised domain adaptation by backpropagation.
\newblock In \emph{ICML}, 1180--1189.

\bibitem[{Graves et~al.(2006)Graves, Fern{\'a}ndez, Gomez, and
  Schmidhuber}]{graves2006connectionist}
Graves, A.; Fern{\'a}ndez, S.; Gomez, F.; and Schmidhuber, J. 2006.
\newblock Connectionist temporal classification.
\newblock In \emph{ICML}.

\bibitem[{Han et~al.(2020)Han, Zhang, Zhang, Yu, Chiu, Qin, Gulati, Pang, and
  Wu}]{han2020contextnet}
Han, W.; Zhang, Z.; Zhang, Y.; Yu, J.; Chiu, C.-C.; Qin, J.; Gulati, A.; Pang,
  R.; and Wu, Y. 2020.
\newblock Contextnet.
\newblock \emph{arXiv:2005.03191}.

\bibitem[{Jain and Jyothi(2018)}]{jain2018improved}
Jain, A.; and Jyothi, P. 2018.
\newblock Improved Accented Speech Recognition Using Accent Embeddings and
  Multi-task Learning.
\newblock In \emph{Interspeech}, 2454--2458.

\bibitem[{Jaiswal et~al.(2020)Jaiswal, Moyer, Ver~Steeg, AbdAlmageed, and
  Natarajan}]{jaiswal2020invariant}
Jaiswal, A.; Moyer, D.; Ver~Steeg, G.; AbdAlmageed, W.; and Natarajan, P. 2020.
\newblock Invariant Representations through Adversarial Forgetting.
\newblock In \emph{AAAI}, 4272--4279.

\bibitem[{Li et~al.(2019)Li, Lavrukhin, Ginsburg, Leary, Kuchaiev, Cohen,
  Nguyen, and Gadde}]{li2019jasper}
Li, J.; Lavrukhin, V.; Ginsburg, B.; Leary, R.; Kuchaiev, O.; Cohen, J.~M.;
  Nguyen, H.; and Gadde, R.~T. 2019.
\newblock Jasper.
\newblock arXiv:1904.03288.

\end{thebibliography}

\section{supplementary material}
In the paper we make a claim that our proposed approach reduces the training time: The reason for this is because in the previous work by \cite{jaiswal2020invariant}, the authors propose to use the raw input as an input (X) to the forget net (F) similarly to the encoder. Which forces the forget net to re-learn the low level features. We argue that it is a redundant step and we should avoid it. Therefore in our proposed method the input to the forget net is the output of the encoder. Suppose the input is X such that the output of the encoder and forget net is Z and M respectively. In our method the input to forget net is Z instead of X. Therefore bypassing the redundant step of re-learning the low level features. It is because of this we argue that our method reduces the training time.

Why do we need global context in the forget net module: Since the forget net has just not to learn what to forget (0s)  but also what to retain (1s). If any of this requires information which is spread out throughout the input we would need global context information to model the same.

The mask values are learned in an adversarial manner. That is why we use the word adversarial forgetting (AF).

\end{document}